\makeatletter
\input{filecontents.sty}
\makeatother

\documentclass[%
titlepage,
secnumarabic,
showpacs,
twocolumn,
nofootinbib,
twoside,
floatfix,
a4paper,
aps]{oxprepr}

\def\Preprint{}

\makeatletter
\ifx\@undefined\Preprint
  \def\JournalPreprint#1#2{#1}
\else
  \def\JournalPreprint#1#2{#2}
\fi
\makeatother

\JournalPreprint{\documentclass[twocolumn,pre,showpacs,floatfix]{revtex4}}{}

\usepackage[american]{babel}
\usepackage{amsfonts}
\usepackage{amsmath}
\usepackage{amssymb}
\usepackage{graphicx}
\usepackage[pdftitle={Local attachment in networks under churn},
pdfauthor={Heiko Bauke and David Sherrington},
pdfcreator={},
pdfsubject={complex networks},
pdfproducer={},
pdfkeywords={local attachment, preferential attachment, complex networks,
  peer-to-peer networks},
hyperfootnotes=false,naturalnames]{hyperref}
\graphicspath{{figures/}}

\IfFileExists{date.tex}{%

}

\newcommand*{\Figurename}{Figure}

\newcommand*{\e}{\mathrm{e}}
\newcommand*{\D}{\mathrm{d}}
\newcommand*{\bigO}[1]{\mathcal{O}\left(#1\right)}

\begin{document}

\title{Local attachment in networks under churn}

\author{Heiko Bauke}
\email[E-mail:]{heiko.bauke@physics.ox.ac.uk}
\author{David Sherrington}
\affiliation{Rudolf Peierls Centre for Theoretical Physics, University of Oxford, 1~Keble~Road, Oxford, OX1~3NP, United Kingdom}

\date{\today}

\begin{abstract}
  In this contribution we introduce \emph{local attachment} as an universal
  network-joining protocol for peer-to-peer networks, social networks, or
  other kinds of networks. Based on this protocol nodes in a finite-size
  network dynamically create power-law connectivity distributions. Nodes or
  peers maintain them in a self-organized statistical way by incorporating
  local information only.
  We investigate the structural and macroscopic properties of such local
  attachment networks by extensive numerical simulations, including
  correlations and scaling relations between exponents. The emergence of the
  power-law degree distribution is further investigated by considering
  preferential attachment with a nonlinear attractiveness function as an
  approximative model for local attachment.
  This study suggests the local attachment scheme as a procedure to be
  included in future peer-to-peer protocols to enable the efficient production
  of stable network topologies in a continuously changing environment.
\end{abstract}


\maketitle

\section{Introduction}
\label{sec:intro}

Information networks and other kinds of networks are an ubiquitous element of
modern infrastructures. These networks can be established by actual hardware
installations as network cables and power lines or may be of conceptual notion
only, as for example in the global network of web pages or in \emph{ad hoc}
peer-to-peer networks.

Networks that are found in nature or in technical or socio-economic systems
differ in their statistical properties remarkably from random networks of
Erd\H{o}s-R{\'e}nyi type. Therefore, much attention has been paid to trying to gain
a better understanding of these networks
\cite{Dorogovtsev:Mendes:2003:Evolution_of_networks,Newman:2004:complex_networks,Newman:Barabasi:Watts:2006:Networks}.

In this paper we discuss a network model which has been inspired by \emph{ad
  hoc} peer-to-peer networks
\cite{Steinmetz:Wehrle:2005:P2P,Lua:etal:2005:Peer-to-Peer}. These kinds of
networks provide an approach to computer connectivity that is complementary to
traditional client-server-architectures. Peer-to-peer networks are designed to
establish a reliable service in a self-organized way without any dependence on
a central instance. Their structural and dynamic properties emerge solely from
local decisions of individual peers.  The lack of a single point of failure
makes peer-to-peer systems potentially more robust than client-server systems,
especially in unreliable constantly changing environments. The peer-to-peer
paradigm became popular through file sharing systems such as Napster or
Gnutella. But file sharing is just the tip of the iceberg, new peer-to-peer
applications such as internet telephony, distributed computing, and
applications in mobile environments are an active field of research
\cite{Steinmetz:Wehrle:2005:P2P}.

In an empirical study of the Gnutella network
\cite{Ripeanu:Iamnitchi:Foster:2002:Gnutella} it has been observed that
peer-to-peer networks have a nontrivial degree distribution in form of a
power-law. In fact, networks with power-law degree distribution $p(k)\sim
k^{-\gamma}$ have properties which might be useful in some peer-to-peer
applications. In power-law networks one can apply local search strategies that
scale sub-linearly with the number of nodes $N$
\cite{Adamic:Lukose:Huberman:2003:Local_Search,Sarshar:etal:2004:Percolation_search},
whereas the search time scales linearly with the system size in networks of
Erd\H{o}s-R{\'e}nyi type. If the exponent of the degree distribution is in the
range $2<\gamma<3$, the diameter\footnote{The diameter if a network refers to
  the longest shortest path between any two nodes in a network.} grows very
slowly, namely proportionally to $\ln\ln N$
\cite{Cohen:Havlin:2003:Ultrasmall}. Furthermore, these networks are very
robust against random node failures
\cite{Albert:etal:2000:Error_tolerance,Cohen:etal:2000:Resilience}.

Several growth models have been proposed, that are able to produce networks
with a power-law degree distribution
\mbox{\cite{Dorogovtsev:Mendes:2003:Evolution_of_networks,Newman:2004:complex_networks,Newman:Barabasi:Watts:2006:Networks}}. On
the other hand, in peer-to-peer networks nodes disappear from the network and
others (or the same later) join the network regularly and over long time
scales peer-to-peer networks have essentially a constant size.  In
straightforward generalizations of growth models to constant size network
models the power-law degree distribution is not necessarily preserved. In
fact, the classical growth model of preferential attachment by Barab{\'a}si and
Albert \cite{Barabasi:Albert:1999:Scaling_in_Random_Networks} fails to produce
power-law networks if nodes enter and disappear at the same rate, see
\cite{Sarshar:Roychowdhury:2004:complex_ad_hoc_networks} and
section~\ref{sec:macroscopic}.

Only a few network evolution models have been proposed, that are able to
produce power-law networks of fixed-size. For example in
\cite{Sarshar:Roychowdhury:2004:complex_ad_hoc_networks} an active rewiring is
introduced to recover the power-law of preferential attachment under node
deletion. The non-growth model \cite{Laird:Jensen:2006:non-growth} gives an
unusual small exponent $\gamma=1$, by incorporating random node deletion and a
non-local node copy operation.

Here we introduce a network model, called \emph{local attachment} model. In
this model node attachment is a genuine local operation. It is able to emulate
peer-to-peer practice and also to maintain a power-law degree distribution for
both growing networks and evolving networks of constant size. Our local
attachment scheme can be used to implement new or improve existing
(unstructured) peer-to-peer protocols.

This paper is organized as follows: In section~\ref{sec:Local_attachment} we
introduce the local attachment mechanism as a network-joining protocol. It
might be utilized to organize so-called constant size networks under churn,
where nodes leave and join a network constantly, or in the context of a pure
growth model. Section~\ref{sec:Local_attachment_churn} investigates local
attachment in constant size networks, while
section~\ref{sec:Local_attachment_growth} focuses on a local attachment growth
model.

\section{The local attachment mechanism}
\label{sec:Local_attachment}

The mechanism that generates the first connection of a new node to an existing
peer-to-peer network (\emph{bootstrapping}) is a crucial detail of the design
of a peer-to-peer protocol. Once a single connection to the network has been
established, the addition of further connections becomes easy. Remarkably,
Gnutella, probably the best documented peer-to-peer protocol
\cite{GPD,Steinmetz:Wehrle:2005:P2P}, does not define the bootstrap process in
detail.

One of the different bootstrap procedures that are actually implemented by
existing software works as follows: It is assumed that a peer who wants to
connect to the network knows already another peer who is likely a member of
the network at this moment. He might remember this peer from connections in
the past or become aware of such a peer by means of a directory service that
knows some members of the network. In the case of Gnutella this service is
called \emph{GWebCache}. After a new peer has got its first connection to the
network he might ask his neighbor about its neighbors. In the jargon of the
Gnutella protocol this is called \emph{pong caching}. The new node may
establish connections to these peers as well, until he owns a reasonable
number of connections.

Real world networks cannot grow forever, after a certain time of growth they
maintain a finite size. In fact, a crucial feature of nodes in peer-to-peer
networks is that they join the network only for a certain time, which is much
shorter than the lifetime of the network. If nodes can leave the network the
process is characterized as \emph{under churn}. Nodes must enter and leave the
network at the same rate, at least on average. Otherwise the network would die
out or grow \emph{ad infinitum}. The rate at which nodes leave (and join) the
network is referred to as the \emph{churn rate}.

This leads us to our model of local attachment of networks under
churn. Initially an arbitrary network of $N$ nodes is given. At each time step
$t$ a randomly chosen node and all edges incident to this node are deleted
from the network and a new node $u=t$ enters the network by introducing $m$
connections to targets in the existing network. Selecting a target from the
current network by local attachment is a two-step procedure. 
\begin{enumerate}
\item First we pick a still-remaining old node $v_1$ randomly with uniform
  probability from the set of nodes with non-zero degree. The new node will
  not be connected to this node; to do so would lead to an exponential degree
  distribution
  \cite{Barabasi:Albert:1999:Scaling_in_Random_Networks,Dorogovtsev:Mendes:2002:Evolution_of_networks}.
\item Instead we assume that the new node $u$ is able to explore the
  neighborhood of $v_1$ and will connect itself to a node $v_2$ uniformly
  randomly chosen among the neighbors of $v_1$.
\end{enumerate}
This two-step procedure is iterated until the degree
of the new node equals $m$. Note that a new node incorporates only minute
information about the network topology to establish its connections; it is a
genuine local process.

If nodes leave the network at a lower rate than others enter the network or
nodes do not leave at all, local attachment leads to a growth model.  In the
growth version of local attachment at each time step $t$ a new node $u=t$
enters the network by connecting to $m$ existing nodes. Each connection is
established by the two-step procedure described above.

Similar approaches to growth models where the attachment of new nodes to the
network is based on local rules have been considered by
\cite{Vazquez:2003:Growing_network,Dorogovtsev:Mendes:Samukhin:2001:Size-dependent_degree_distribution,Krapivsky:Redner:2001:growing_networks}.
In the random walk model \cite{Vazquez:2003:Growing_network} new nodes are
added to the network by establishing a directed edge to some randomly chosen
node. This model resembles a random walk on the network because each time when
an edge is created to a vertex in the network then with some probability $p$
an edge is also created to one of the nearest neighbors of this vertex. The
walk stops with probability $(1-p)$ and the next new node is added to the
network. In the model considered in
\cite{Dorogovtsev:Mendes:Samukhin:2001:Size-dependent_degree_distribution} new
nodes are connected to both ends of a randomly chosen edge by two undirected
links. This leads to networks with very high clustering.  In
\cite{Krapivsky:Redner:2001:growing_networks} the authors consider a growth
process of directed networks which results in networks with a simple treelike
topology. At each time each node has an out-degree of one but an arbitrary
in-degree. In this model each new node $u$ is either connected to a randomly
chosen node $v$ or to the unique neighbor of $v$. Both processes happen with
some probability $(1-p)$ and $p$, respectively. A similar model has been
introduced in the context of the world wide web
\cite{Kleinberg:etal:1999:Web}.

Although local attachment has been motivated in the context of peer-to-peer
networks, it may model other kinds of networks as well. Social networks where
old members introduce new members to their neighbors are an obvious example.

\section{Local attachment networks under churn}
\label{sec:Local_attachment_churn}

\subsection{Macroscopic quantities}
\label{sec:macroscopic}

If nodes enter and leave a network at the same rate, the microscopic structure
of the network is constantly changing. But there are macroscopic quantities
that evolve asymptotically to limiting values (in a statistical sense),
e.\,g.\ the degree distribution or the number of edges. If old nodes leave
randomly and new nodes enter the network by making $m$ connections to the
existing network, the number of edges in the network will fluctuate around $m
N/2$ independently of the details of the underlying microscopic dynamics of
the network-joining protocol. But in general, other macroscopic quantities
depend on the details of the way that new nodes connect to the network.

Barab{\'a}si and Albert introduced in
\cite{Barabasi:Albert:1999:Scaling_in_Random_Networks} the mechanism of
\emph{preferential attachment}, which yields in growing networks (no nodes are
removed) a power-law degree distribution \mbox{$p(k)\sim k^{-\gamma}$} with
exponent $\gamma=3$.  The attachment mechanism assigns explicitly to each node
an attractiveness $A(k)$ that is proportional to its degree $k$. Whenever a
new node $u$ is added to the growing network it attaches to $m$ old nodes $v$
which are chosen randomly with probability
\begin{equation}
  p_\mathrm{add}(v) = \frac{A(k_v)}{\sum_iA(k_i)} = \frac{k_v}{2mt}\,,
\end{equation}
where $k_i$ is the degree of node $i$ and $t=N$ is the current number of nodes
in the network.  This growth process starts from some small initially-given
network, e.\,g.\ $m+1$ fully connected nodes, and the quantity $t$ may be
interpreted as the time that passed since the growth process started.

\begin{figure}
  \includegraphics[width=\linewidth]{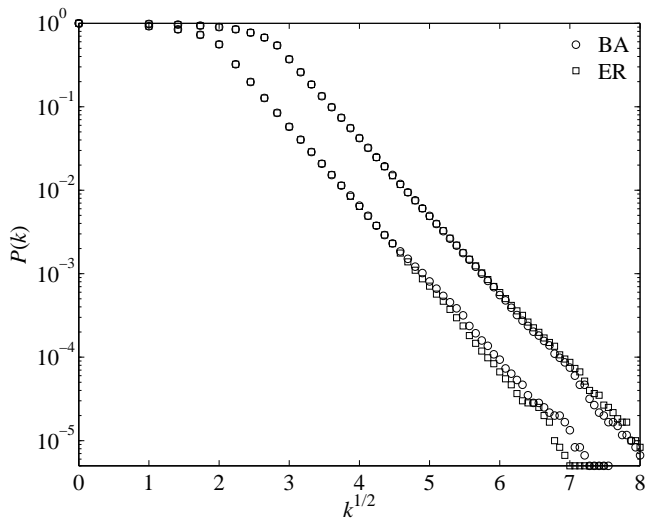}
  \caption{Cumulative degree distribution $P(k)=\sum_{i\ge k} p(i)$ for
    evolving networks of constant size undergoing preferential attachment
    under churn with $N=20\,000$ nodes and a mean degree $m$.  Randomly chosen
    nodes disappear from the network and new nodes enter the network at the
    same rate. New nodes connect to $m$ old nodes via preferential attachment
    (lower line $m=4$, upper line $m=8$). Initial networks were given either
    by an Erd\H{o}s-R{\'e}nyi (ER) network or by a Barab{\'a}si-Albert (BA) network.}
  \label{fig:Barabasi_Albert_model_dyn}
\end{figure}
\begin{figure}
  \includegraphics[width=\linewidth]{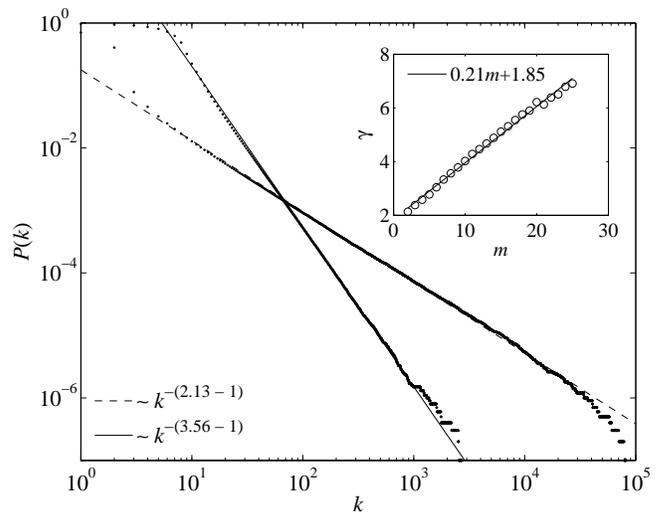}
  \caption{Cumulative degree distribution $P(k)=\sum_{i\ge k} p(i)$ for a
    network under churn evolving via local attachment and a fit of this
    distribution to a power-law $P(k)\sim k^{-(\gamma-1)}$, which corresponds
    to $p(k)\sim k^{-\gamma}$. Randomly chosen nodes disappear from the
    network and new nodes enter the network at the same rate. New nodes
    connect to $m$ old nodes via local attachment.  In these simulations the
    initial networks were Erd\H{o}s-R{\'e}nyi with $mN/2$ edges, but the
    steady-state power-law behavior is independent of the details of the
    starting state.  Results in the main figure are shown for $m=2$ (dashed
    line), and $m=8$ (solid line) for $N=100\,000$ nodes.  In the inset:
    Exponent $\gamma$ of the power-law degree distribution $p(k)$ as a
    function of $m$.  Distributions and exponents have been determined by
    averaging over 100 networks.}
  \label{fig:local_attachment_dyn_gamma}
\end{figure}
\begin{figure}
  \includegraphics[width=\linewidth]{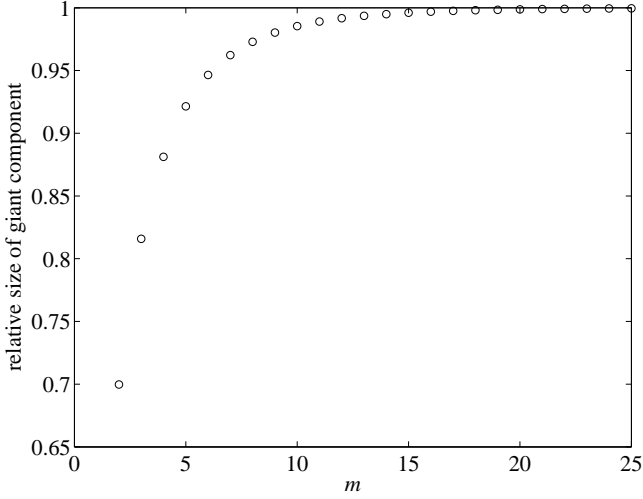}
  \caption{Mean size of the largest connected component of local attachment
    networks under churn of $N=100\,000$ nodes but with different mean
    degree~$m$.}
  \label{fig:local_attachment_dyn_gc}
\end{figure}

However, the power-law degree distribution of the Bara\-b{\'a}si-Albert model is
not preserved if the network is under churn.  If new nodes enter the network
via preferential attachment but random nodes leave the network at the same
rate \cite{Sarshar:Roychowdhury:2004:complex_ad_hoc_networks}, the degree
distribution converges independently of the initial network to a distribution
with a tail that falls faster than any power-law. In fact, we find numerically
that the tail of its cumulative degree distribution $P(k)=\sum_{i\ge k} p(i)$
falls as $\bigO{\exp(-{k^{1/2}})}$ and therefore
$p(k)\sim\bigO{\exp(-{k^{1/2}})}$, see
\figurename~\ref{fig:Barabasi_Albert_model_dyn}. To maintain the power-law
degree distribution under node deletion a deletion-compensation protocol was
introduced by \cite{Sarshar:Roychowdhury:2004:complex_ad_hoc_networks} in
which, if a node has lost a connection, it initiates on average $n$ new
connections by choosing nodes by preferential attachment. This allows for any
given deletion rate to tune the power-law exponent of the degree distribution
to be anywhere in $(2,\infty)$ by varying the average number of compensatory
edges for each deleted edge $n$.

On the contrary, if new nodes enter the network by local attachment instead of
preferential attachment, the deletion-compensation protocol becomes
unnecessary. In fact, if new nodes are inserted via local attachment, the
degree distribution exhibits a power-law even if nodes leave and enter at the
same rate, see \figurename~\ref{fig:local_attachment_dyn_gamma}. Numerically
we find that the exponent $\gamma$ of the degree distribution $p(k)$ grows
linearly with the mean degree $m$ of the network. This allows one to tune the
exponent $\gamma$ by choosing an appropriate mean degree $m$.

For a peer-to-peer network it is desirable that the network stays connected,
even if nodes leave (and enter) the network constantly. Of course, the more
edges a network contains (parameter $m$ large), the smaller the probability
that the network falls apart into more than one connected
component. Numerically we find that networks that evolve under the proposed
dynamics consist of a large connected component and isolated nodes or small
components of size $\bigO{1}$, see
\figurename~\ref{fig:local_attachment_dyn_gc}. The number of nodes outside the
largest connected component is rather small compared to the size of the
largest component. If the mean degree $m$ is greater than or equal to five,
less than 10\,\% of the nodes are not part of the large component.

\subsection{Node attractiveness and degree correlations}

If a new node $u$ establishes an edge to the network by our local attachment
rule, the connection will be made to node $v_2$ with probability
\begin{equation}
  \label{eq:p_add}
  p_\mathrm{add}(v_2) = \frac{1}{N} \sum_{v_1\in\{\text{nn } v_2\}} \frac{1}{k_{v_1}} \,,
\end{equation}
where the sum runs over the neighbors of node $v_2$ and $k_{v_1}$ denotes the
degree of node $v_1$. The probability (\ref{eq:p_add}) can be rephrased in
terms of a node attractiveness $A(v_2)$ by
\begin{equation}
  p_\mathrm{add}(v_2) = \frac{A(v_2)}{\displaystyle\sum_w A(w)}
  \quad\text{with}\quad
  A(v_2) \sim \sum_{v_1\in\{\text{nn } v_2\}} \frac{1}{k_{v_1}} \,.
\end{equation}
Note that in general the attractiveness of a node is determined uniquely up to
a multiplicative constant only. Because of the two-step nature of local
attachment the attractiveness of a node depends on its own degree and the
degrees of its nearest neighbors, too, and therefore correlations between the
degrees of neighboring nodes are important.

In uncorrelated networks local attachment is equivalent to preferential
attachment, in the sense that a node with degree $k$ has \emph{on average} an
attractiveness proportional to its degree. Let $\langle A(k)\rangle$ be the
average attractiveness of a node with degree $k$ and $p_\mathrm{nn}(i|k)$ the
probability that the degree of a nearest neighbor of a node with degree $k$
equals $i$, then
\begin{equation}
  \langle A(k)\rangle = 
  \frac{\displaystyle\sum_{i_1,\dots,i_k=1}^\infty
    \left(\sum_{j=1}^k\frac{1}{i_j} \right)
    \prod_{h=1}^k p_\mathrm{nn}(i_h|k)
  }
  {\displaystyle\sum_{i_1,\dots,i_k=1}^\infty\prod_{h=1}^k p_\mathrm{nn}(i_h|k)} \,,
\end{equation}
where $i_h$ is the degree of the $h$th neighbor of a node with degree $k$.

\JournalPreprint{}{\enlargethispage{\baselineskip}}%
For uncorrelated networks $p_\mathrm{nn}(i|k)$ does not depend on $k$ and
equals the ordinary degree distribution $p(i)$ and thus we get
\begin{equation}
  \label{eq:A_k_lin}
  \langle A(k)\rangle = k \frac{\sum_{i=1}^\infty p(i)/i}{1-p(0)}\,.
\end{equation}
After an arbitrary rescaling (\ref{eq:A_k_lin}) equals the attractiveness of
the Barab{\'a}si-Albert model. But note that in Barab{\'a}si-Albert networks the
attractiveness $A(k)\sim k$ is imposed and these networks have structural
properties that are different from uncorrelated networks.

\begin{figure}
  \includegraphics[width=\linewidth]{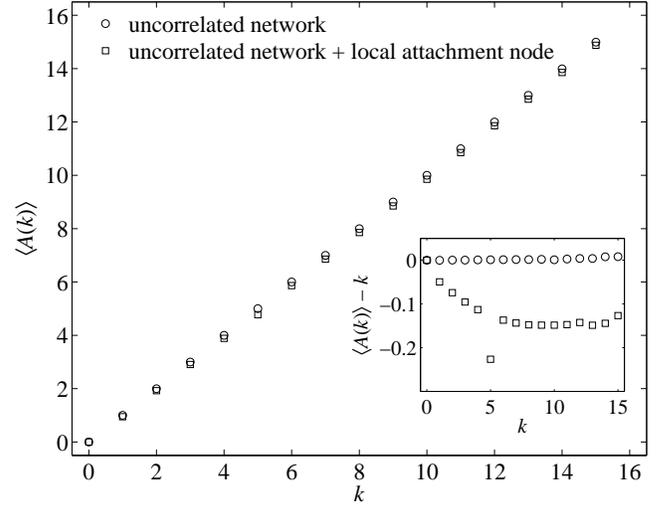}
  \caption{Average attractiveness $\langle A(k)\rangle$ as a function of the
    degree for uncorrelated networks and for networks in which a single extra
    node has been added to an uncorrelated network via $m=5$ new edges. The
    uncorrelated networks have $N=40$ nodes and each edge is present with
    probability $m/N$ independently. Results have been averaged over $10^6$
    networks.}
  \label{fig:ER_local_attachment_dyn}
\end{figure}
\begin{figure}
  \includegraphics[width=\linewidth]{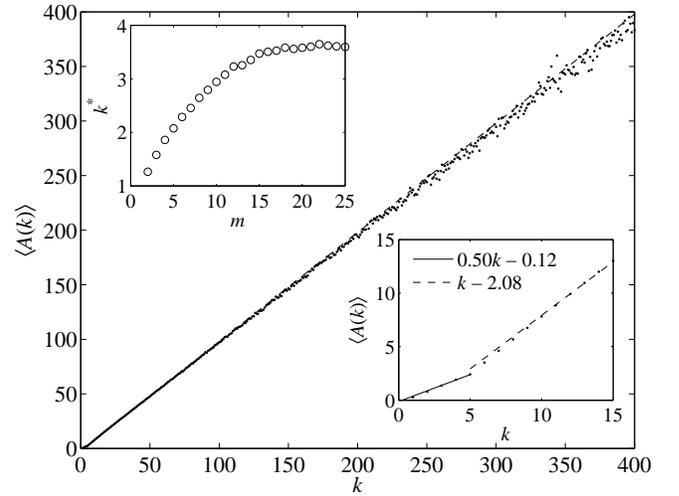}
  \caption{Mean attractiveness of a node as a function of its degree in a
    local attachment network under churn for $m=5$.  The inset in the right
    lower corner shows a detail of this distribution for small degrees. After
    an arbitrary rescaling the attractiveness $\langle A(k)\rangle$ is
    approximately given by $\langle A(k)\rangle\approx k-k^*$ for degrees
    $k\ge m$. The offset $k^*$ as a function of the mean degree of the network
    is shown in the inset in the upper left.}
  \label{fig:local_attachment_dyn_attractiveness}
\end{figure}

On the other hand, the local attachment mechanism \emph{does} induce
correlations and as a consequence the average attractiveness $\langle
A(k)\rangle$ does not follow the linear law~(\ref{eq:A_k_lin}). This can be
illustrated by a simple numerical experiment. 

In this experiment a single node enters an uncorrelated Erd\H{o}s-R{\'e}nyi
network by the local attachment rule. For each network the attractiveness as a
function of the degree has been determined \emph{before} node attachment and
\emph{afterward}. The attractiveness was averaged over a large number of
networks and has been normalized arbitrarily such that $\langle A(k)\rangle$
grows with a slope of one for $k>m$, see
\figurename~\ref{fig:ER_local_attachment_dyn}.  The effect of the correlations
induced by a single local attachment node is minute but clear, $\langle
A(k)\rangle$ becomes a nonlinear function. In fact, it is piecewise linear
with different slope for $k<m$ and $k>m$.

The more nodes enter the network by local attachment the stronger the
degree-degree-correlations, but on the other hand random node removal destroys
correlations. The competition between these two processes leads finally to a
dynamical steady state. Empirically we find that in this steady state the
average attractiveness $\langle A(k)\rangle$ is (after an arbitrary
normalization) approximately given by the piecewise linear function
\begin{equation}
  \label{eq:mean_A}
  \langle A(k)\rangle =
  \begin{cases}
    \left(1-\dfrac{k^*}{m}\right)k & \text{if $0\le k \le m$} \\[1.75ex]
    k-k^* & \text{if $k \ge m$} \,,
  \end{cases}
\end{equation}
see \figurename~\ref{fig:local_attachment_dyn_attractiveness}.

Other quantities that are sensitive to degree-degree-correl\-ations are the
distribution
\begin{equation}
  \label{eq:k_nn}
  \langle k_\mathrm{nn}(k)\rangle =
  \sum_i i  p_\mathrm{nn}(i|k)
\end{equation}
of the mean degree of the neighboring nodes of a node with degree $k$ and the
distribution 
\begin{equation}
  \label{eq:Delta_k_nn}
  \Delta{k}_\mathrm{nn}(k) = \sqrt{
    \sum_i \left(
      i - \langle k_\mathrm{nn}(k)\rangle
    \right)^2 p_\mathrm{nn}(i|k) 
  }
\end{equation}
of the standard deviation of the degree distribution of the neighboring nodes
of a node. In uncorrelated networks both quantities do not depend on the
degree $k$. 

\begin{figure}
  \includegraphics[width=\linewidth]{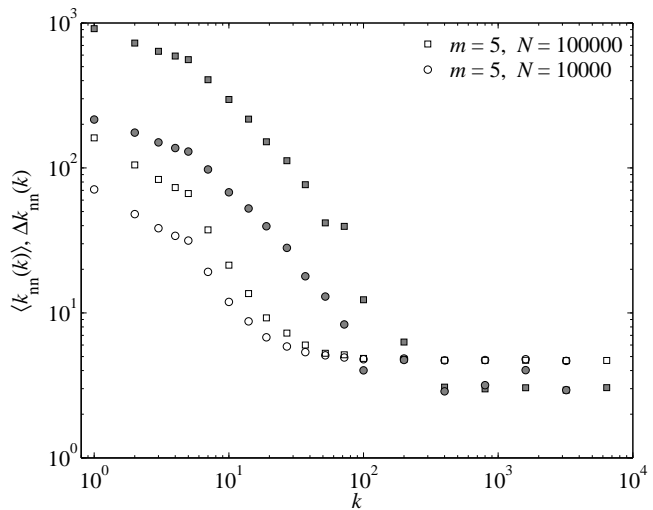}
  \caption{Distribution $\langle k_\mathrm{nn}(k)\rangle$ of the mean degree
    of the neighboring nodes (open symbols) and the distribution
    $\Delta{k}_\mathrm{nn}(k)$ of the standard deviation of the degree
    distribution of the neighboring nodes (closed symbols) of a node with
    degree $k$ for nodes in local attachment networks under churn.}
  \label{fig:local_attachment_dyn_P_nn_k}
\end{figure}

However, our local attachment mechanism does induce correlations between low
degree nodes and their neighbors and the distributions (\ref{eq:k_nn}) and
(\ref{eq:Delta_k_nn}) are not flat. As a consequence of the constant deletion
and addition of nodes the distributions (\ref{eq:k_nn}) and
(\ref{eq:Delta_k_nn}) of local attachment networks under churn can be divided
into two different regimes. At the level of large $k$
degree-degree-correlations are averaged out and the distribution $\langle
k_\mathrm{nn}(k)\rangle$ is almost constant and roughly equals the mean degree
of the whole network $m$, independently of the system size $N$, see
\figurename~\ref{fig:local_attachment_dyn_P_nn_k}. Furthermore, the
distribution of the degrees of neighboring nodes is rather narrow and
therefore $\Delta{k}_\mathrm{nn}(k)<\langle k_\mathrm{nn}(k)\rangle$. On the
other hand, low degree nodes attach preferably to high degree nodes and the
degree $\langle k_\mathrm{nn}(k)\rangle$ is much larger than the mean degree
of the whole network $m$. One says, the network shows \emph{disassortative}
correlations. The distribution of the degrees of neighboring nodes is very
broad in this regime and at fixed $k$ both quantities $\langle
k_\mathrm{nn}(k)\rangle$ and $\Delta{k}_\mathrm{nn}(k)$ grow with the system
size $N$.

\subsection{The degree distribution: theoretical considerations}

In the last section we have shown that correlations are an important feature
of local attachment networks under churn. A full mathematical description of
these networks has to take them into account. This is a nontrivial
task. However, the power-law degree distribution of local attachment networks
under churn can be explained qualitatively by a self-consistent description,
if only correlations on the level of the nonlinear mean attractiveness
(\ref{eq:mean_A}) are taken into account.

\subsubsection{Preferential attachment with the nonlinear attractiveness}
\label{sec:Preferential_attachment}

For this description we consider a preferential attachment model under churn
with the nonlinear attractiveness
\begin{equation}
  \label{eq:A_nonlin}
  A(k) =
  \begin{cases}
    \left(1-\dfrac{k^*}{m}\right)k & \text{if $0\le k \le m$} \\[1.75ex]
    k-k^* & \text{if $k \ge m$} \,,
  \end{cases}
\end{equation}
as an approximative model for our local attachment model. Note that in this
nonlinear preferential attachment model the attractiveness (\ref{eq:A_nonlin})
is imposed explicitly, while in the local attachment model the nonlinear mean
attractiveness (\ref{eq:mean_A}) emerges implicitly from the dynamics of the
model.

Let $D(i, t)$ be the probability that a node that entered the network at time
$i$ is still present at time $t\ge i$. Node removal does not depend on time
$t$ nor on the removal of other nodes and therefore $D(i, t)$ depends only on
the age $t'=t-i$ of a node. For $N\to\infty$ and $(t-i)\to \infty$ such that
$(t-i)/N\to\text{const.}$ the probability $D(i, t)$ becomes
\begin{equation}
  \label{eq:prob_in}
  D(i, t) = D(t') = \left(1-\dfrac{1}{N}\right)^{t'} \approx
  \e^{-\frac{t'}{N}} \,.
\end{equation}

For the calculation of the degree distribution $p(k)$ we adopt a continuous
variable approximation as used in
\cite{Dorogovtsev:Mendes:2001:Scaling_properties,Sarshar:Roychowdhury:2004:complex_ad_hoc_networks}. Let
$\langle k(i, t)\rangle$ be the mean degree (ensemble average over different
network realizations) at time $t\ge i$ of a node that has entered the network
at time $i$ and has not yet disappeared from the network. In the framework of
a continuous approximation the evolution of $\langle k(i, t)\rangle$ is given
by
\begin{equation}
  \label{eq:evolution}
  \frac{\partial \langle k(i, t)\rangle}{\partial t} =
  m \frac{A(\langle k(i, t)\rangle)}{S(t)} - 
  \frac{\langle k(i, t)\rangle}{N} 
\end{equation}
with the initial condition $\langle k(i, i)\rangle=m$, where the quantity
$S(t)$ is given by
\begin{equation}
  S(t) = 
  \int_0^t D(i, t) A(\langle k(i, t)\rangle) \,\D i\,.
\end{equation}

\begin{figure*}
  \includegraphics[width=0.482\linewidth]{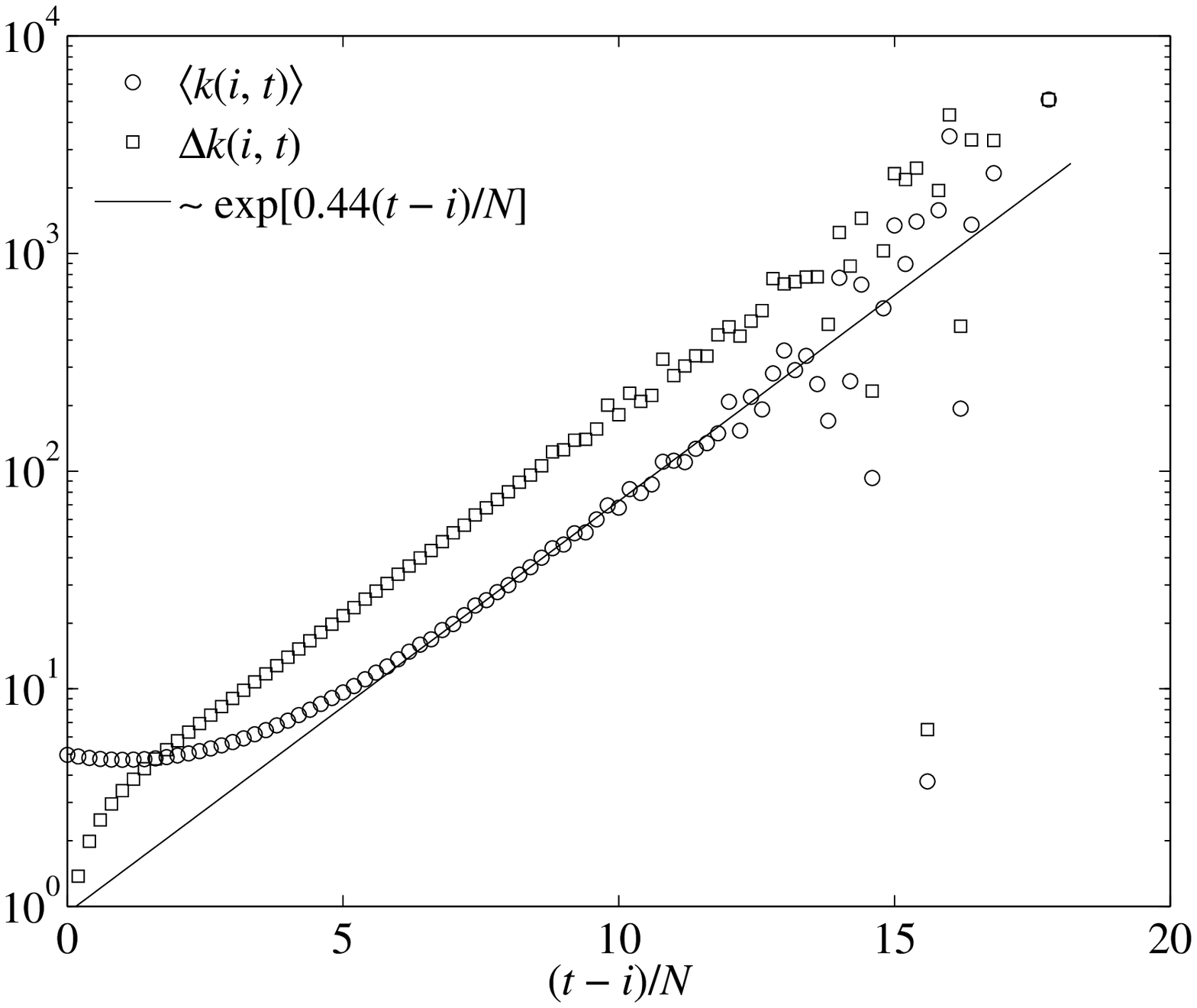}
  \hfill
  \includegraphics[width=0.482\linewidth]{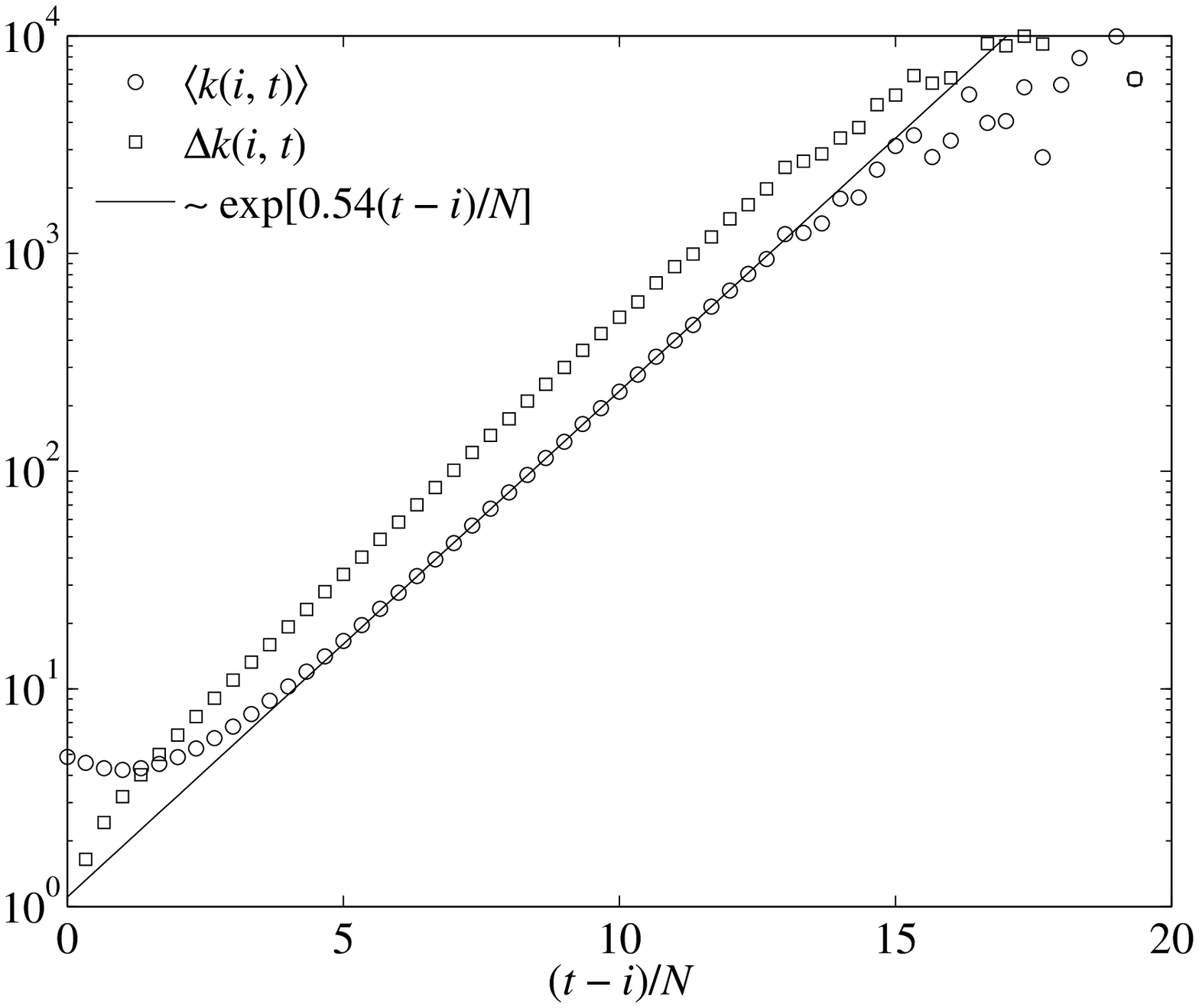}
  \caption{Mean degree $\langle k(i, t)\rangle$ of a node and the standard
    deviation $\Delta k(i,t)$ of the distribution of the degree as a function
    of its age $(t-i)$. The diagrams show results for two different kinds of
    networks under churn, a preferential attachment network with a nonlinear
    attractiveness \eqref{eq:A_nonlin} (left) and a local attachment network
    (right). Both networks have $N=100\,000$ nodes, a mean degree $m=5$, and
    the parameter of the attractiveness function equals $k^*=2.077$.}
  \label{fig:local_attachment_dyn_growth}
\end{figure*}

If the degree distribution $p(k)$ of a fixed-size network under churn
converges for $t\to \infty$ to a limiting distribution, then $S(t)$ must
converge to a constant. This motivates us to introduce the quantity
\begin{align}
  s & = \lim_{t\to\infty}\frac{S(t)}{N} = \frac{1}{N} \lim_{t\to\infty}
  \int_0^t D(i, t) A(\langle k(i, t)\rangle)\,\D i\,.\\
  \intertext{After the network has reached some time-independent
    degree-distribution $s$ is given by the mean attractiveness of the
    network}
  s & = \int_0^\infty p(k) A(k)\,\D k\,.\\
  \intertext{The mean attractiveness $s$ depends on the unknown degree
    distribution $p(k)$, but we can make some general statements about $s$ by
    taking into account the imposed attractiveness $A(k)$. If we assume that
    $A(k)$ is given by (\ref{eq:A_nonlin}) then} s & = \int_0^m p(k)
  \left(1-\dfrac{k^*}{m}\right)k\,\D k +
  \int_m^\infty p(k) (k-k^*)k\,\D k \nonumber\\
  \label{eq:s}
  & = m - k^* \left(
    \int_0^m \frac{k}{m}p(k)\,\D k + \int_m^\infty p(k) \,\D k
  \right)\,,
\end{align}
where $m$ is the mean degree of the network. From (\ref{eq:s}) follows the
inequality $m-k^*<s<m$.

The attractiveness (\ref{eq:A_nonlin}) is piecewise linear, therefore we have
to distinguish two cases for the solution of (\ref{eq:evolution}). For
$\langle k(i, t)\rangle<m$ equation (\ref{eq:evolution}) becomes
\begin{align}
  \frac{\partial \langle k(i, t)\rangle}{\partial t} & =
  \frac{m}{s N}\left(1-\dfrac{k^*}{m}\right)\langle k(i, t)\rangle -
  \frac{1}{N}\langle k(i, t)\rangle \nonumber\\
  & = \frac{m-k^*-s}{s N}\langle k(i, t)\rangle\,.
\end{align}
Because $(m-k^*-s)<0$, this equation has exponentially decaying solutions for
any initial condition $\langle k(i, t_0)\rangle=k_0$ with $t_0\ge i$,
\begin{equation}
  \langle k(i, t)\rangle = k_0 \e^{\frac{m-k^*-s}{s} \frac{t-t_0}{N}} \,.
\end{equation}
On the other hand, for $\langle k(i, t)\rangle>m$ we find the evolution
equation
\begin{align}
  \frac{\partial \langle k(i, t)\rangle}{\partial t} & =
  m\frac{\langle k(i, t)\rangle - k^*}{s N} - 
  \frac{1}{N}\langle k(i, t)\rangle \nonumber\\
  & = \frac{m-s}{s N}\langle k(i, t)\rangle - \frac{m k^*}{s N}\,,
\end{align}
which has the solution
\begin{equation}
  \label{eq:sol2}
  \langle k(i, t)\rangle = \frac{m k^*}{m-s} + \left(
    k_0 - \frac{m k^*}{m-s}
  \right) \e^{\frac{m-s}{s} \frac{t-t_0}{N}}
\end{equation}
for the initial condition $\langle k(i, t_0)\rangle=k_0$ with $t_0\ge i$.  The
exponent $(m-s)/s$ in (\ref{eq:sol2}) is positive, but the prefactor $[k_0-m
k^*/(m-s)]$ might be positive or negative depending on the initial condition,
which corresponds to exponentially growing or shrinking $\langle k(i,
t)\rangle$.

New nodes enter the network by establishing $m$ new edges at time $t=i$ and
thus the solution (\ref{eq:sol2}) reads
\begin{equation}
  \label{eq:k_i_t}
  \langle k(i, t)\rangle = \frac{m k^*}{m-s} + 
  \frac {m(m-k^*-s)}{m-s} \e^{\frac{m-s}{s} \frac{t-i}{N}}\,.
\end{equation}
Because $(m-k^*-s)<0$, this solution predicts an exponentially fast decay of
$\langle k(i, t)\rangle$ for (young) nodes with degree $\gtrsim m$. The
initial condition $\langle k(i, i)\rangle=m$ is just on the borderline, but in
both cases our calculations predict a decaying mean degree $\langle k(i,
t)\rangle$.

\Figurename~\ref{fig:local_attachment_dyn_growth} shows that $\langle k(i,
t)\rangle$ is indeed decaying, but only for young nodes. In fact, if a node
has reached a sufficient large age $\langle k(i, t)\rangle$ grows
exponentially.  The crossover from decaying to exponentially growing $\langle
k(i, t)\rangle$ is driven by statistical fluctuations among the degrees of
individual nodes. Because of these statistical fluctuations individual nodes
can gain some degree $k_0$ such that $[k_0 - m k^*/(m-s)] > 0$. In this case
(\ref{eq:sol2}) predicts that the degree of such nodes will grow exponentially
fast (on average).

This interpretation is compatible with our numerical experiments, as
\figurename~\ref{fig:local_attachment_dyn_growth} shows. By construction,
nodes enter the network by establishing exactly $m$ connections. Then the
degree of young nodes shrinks on average but simultaneously the distribution
of the degree of nodes of the same age gets broader with increasing age. As a
consequence, a certain fraction of nodes can reach a critical degree, such
that exponential growth can take off.

According to (\ref{eq:k_i_t}) the exponent of the exponential growth is given
by
\begin{equation}
  \label{eq:beta}
  \beta = \frac{m-s}{s}\,.
\end{equation}
We may use the degree distribution and $\langle k(i, t)\rangle$ to determine
numerical values for $s$ (see equation (\ref{eq:s})) and $\beta$
independently. Our results confirm that in the case of preferential attachment
networks with a nonlinear attractiveness \eqref{eq:A_nonlin} the relation
(\ref{eq:beta}) is approximately fulfilled, see
\tablename~\ref{tab:parameters_dyn}.

\begin{table}[b]
  \caption{Characteristic quantities of preferential attachment networks with a
    nonlinear attractiveness \eqref{eq:A_nonlin} under churn and local attachment
    networks under churn, see text for an explanation of $m$, $k^*$, $s$,
    $\beta$, and~$\gamma$.} 
  \label{tab:parameters_dyn}
  \begin{ruledtabular}
    \begin{tabular}{@{}cccccccc@{}}
      $m$ & $k^*$ & $s$    & $\beta$ & $(m-s)/s$ & $\gamma$ & $m/(m-s)$ & $1/(\gamma-1)$\\
      \hline
      \multicolumn{8}{c}{\makebox[0pt]{preferential attachment networks with 
          nonlinear attractiveness}}\\
      3	& 1.58 & 1.92 & 0.551 & 0.559 & 2.85 & 2.79 & 0.540 \\
      4	& 1.86 & 2.66 & 0.491 & 0.503 & 3.04 & 2.99 & 0.490 \\
      5	& 2.08 & 3.44 & 0.438 & 0.453 & 3.34 & 3.21 & 0.427 \\
      6	& 2.29 & 4.23 & 0.399 & 0.418 & 3.50 & 3.39 & 0.400 \\[0.67ex]
      \multicolumn{8}{c}{local attachment networks} \\
      3 & 1.58 & 2.02 & 0.686 & 0.488 & 2.38 & 3.05 & 0.725 \\
      4 & 1.86 & 2.74 & 0.607 & 0.459 & 2.58 & 3.18 & 0.633 \\
      5 & 2.08 & 3.51 & 0.535 & 0.425 & 2.78 & 3.35 & 0.562 \\
      6 & 2.29 & 4.28 & 0.478 & 0.400 & 3.04 & 3.50 & 0.490 \\
    \end{tabular}
  \end{ruledtabular}
\end{table}

Note that in the nonlinear preferential attachment model of constant size
networks under churn the mean degree of a node grows exponentially with its
age $t'=t-i$ proportionally to $\e^{\beta t'/N}$, whereas in growth models
that yield a power-law degree distribution this quantity grows proportional to
some power-law $(t/i)^\beta$
\cite{Dorogovtsev:Mendes:Samukhin:2000:Structure_of_Growing_Networks}.

The degree distribution $p(k)$ follows from the mean degree $\langle k(i,
t)\rangle$. Our numerical findings show that for sufficiently large ages
$t'=t-i$ the distribution $\langle k(i, t)\rangle$ is given by an exponential
law
\begin{align}
  \langle k(i, t)\rangle = \langle k(t')\rangle & = 
  c_1 + c_2 \e^{\frac{m-s}{s} \frac{t'}{N}}\,.
\end{align}
The probability $D(t')$ that a node $i$ is in the network at time $t$ falls
exponentially with its age $t'$.  In general, the degree distribution is given
by
\begin{equation}
  p(k) = \frac{1}{N} D(t'_k) \left|
    \frac{\partial \langle k(t')\rangle}{\partial t'}
  \right|_{t'=t'_k}^{-1}\,,
\end{equation}
where $t'_k$ is the solution of the equation $\langle k(t'_k)\rangle=k$ and %
$\left|{\partial\langle k(t')\rangle}/{\partial t'}\right|_{t'=t'_k}^{-1}$ is
approximately the number of nodes with degree $k$ (neglecting node
deletion). With (\ref{eq:prob_in}) we get
\begin{equation}
  \frac{\partial \langle k(t')\rangle}{\partial t'} = 
  \frac{m-s}{s}\frac{\langle k(t')\rangle}{N}
  \quad\text{and}\quad
  \frac{m-s}{s}\frac{t'}{N} = \ln\frac{k-c_1}{c_2}\,.
\end{equation}
Then (assuming $k>c_1$) the degree distribution is given by
\begin{equation}
  p(k) = \frac{s}{m-s}c_2^{\frac{s}{m-s}} (k - c_1)^{-\frac{m}{m-s}}
\end{equation}
and the degree distribution $p(k)$ obeys a power-law tail with the exponent
\begin{equation}
  \label{eq:gamma}
  \gamma=\frac{m}{m-s} \,.
\end{equation}
If the parameters $s$ and $\gamma$ are determined independently from the
degree distribution $p(k)$ by numerical experiments for preferential
attachment networks with a nonlinear attractiveness \eqref{eq:A_nonlin}, we
find a reasonable agreement between $\gamma$ and the predicted exponent
(\ref{eq:gamma}), see \tablename~\ref{tab:parameters_dyn}.

\subsubsection{Local attachment}

Preferential attachment with nonlinear attractiveness was introduced as an
approximative description for our original local attachment model. It takes
into account correlations on the level of the nonlinear mean attractiveness
(\ref{eq:mean_A}) that is observed in local attachment networks under
churn. 

\begin{figure}
  \includegraphics[width=0.8\linewidth]{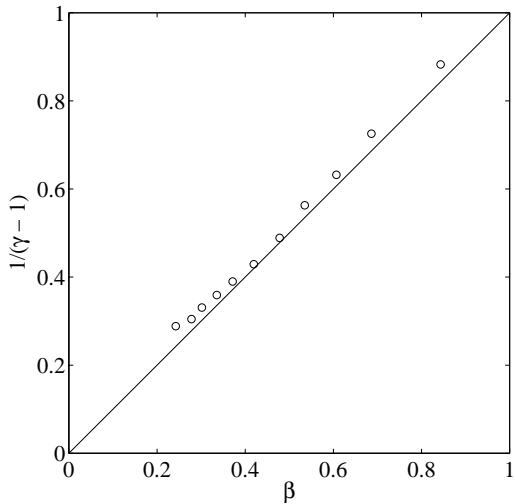}
  \caption{Scaling relation between exponents $\beta$ and $\gamma$ of local
    attachment networks under churn. Simulations have been carried out with
    networks of $N=100\,000$ nodes with mean degree $2\le m \le 12$.}
  \label{fig:local_attachment_dyn_beta_gamma}
\end{figure}

In this way our simplified model is able to cover some important qualitative
features of local attachment networks under churn. It accounts for the
nonlinear mean attractiveness, for the exponential grow of $\langle k(i,
t)\rangle$ (see \figurename~\ref{fig:local_attachment_dyn_growth}), and for
the power-law degree distribution $p(k)$. Moreover, the combination of
(\ref{eq:beta}) and (\ref{eq:gamma}) gives the scaling relation
\begin{equation}
  \label{eq:gamma_beta}
  \beta = \frac{1}{\gamma - 1} 
\end{equation}
for preferential attachment networks under churn with nonlinear
attractiveness. In fact, we find numerically that the universal scaling
relation (\ref{eq:gamma_beta}) is approximately fulfilled for nonlinear
preferential attachment networks under churn as well as for local attachment
networks under churn, see
\figurename~\ref{fig:local_attachment_dyn_beta_gamma} and
\tablename~\ref{tab:parameters_dyn}.

On the other hand, local attachment networks have higher order correlations
that are not taken into account by the simplified model of nonlinear
preferential attachment. As a consequence, not all results of
section~\ref{sec:Preferential_attachment} hold also for our original model of
local attachment networks under churn. For example numerical results show that
the equations (\ref{eq:beta}) and (\ref{eq:gamma}) that relate the mean
attractiveness $s$ and the mean degree $m$ to the exponents $\beta$ and
$\gamma$ are not fulfilled in the case of our original local attachment
networks, see \tablename~\ref{tab:parameters_dyn}.  Local attachment networks
and nonlinear preferential attachment networks with the same mean degree and
the same mean attractiveness have different exponents $\beta$ and $\gamma$.

\section{Growing local attachment networks}
\label{sec:Local_attachment_growth}

In this section we consider a network growth process without churn that
utilizes local attachment as a network-joining protocol as introduced in
section~\ref{sec:Local_attachment}. The growth process starts from some
initially-given network without isolated nodes, e.\,g.\ a network of $m+1$
fully connected nodes or some other small connected network, and constructs a
network that has in the limit of infinite large networks a mean degree $2m$
independently of the initial configuration.

\subsection{Degree distribution}

\begin{figure}
  \includegraphics[width=\linewidth]{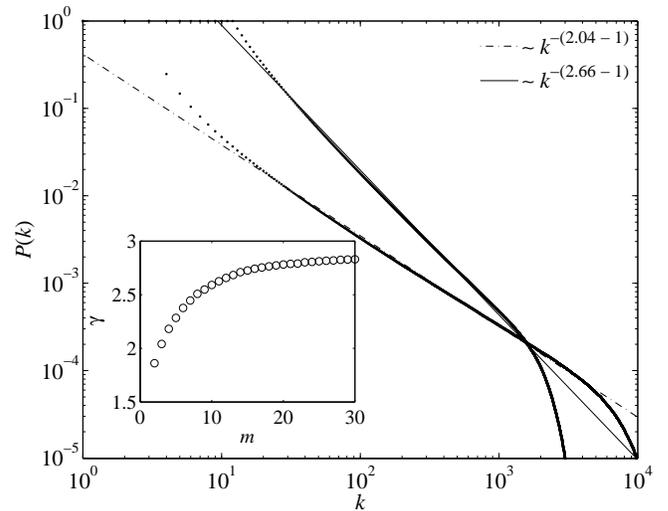}
  \caption{Cumulative degree distribution $P(k)=\sum_{i\ge k} p(i)$ of a
    growing local attachment network and a fit of this distribution to a
    power-law $P(k)\sim k^{-(\gamma-1)}$, which corresponds to $p(k)\sim
    k^{-\gamma}$. In the inset: Exponent $\gamma$ of the power-law degree
    distribution as a function of the number $m$ of edges that connect a new
    node to the network.  Results are obtained for growing networks generated
    by the local attachment rule with $t=N=100\,000$ nodes and $m=3$ (dash-dot
    line) and $m=12$ (dashed line).}
  \label{fig:ba_local_degree_dist}
\end{figure}
\begin{figure}
  \includegraphics[width=\linewidth]{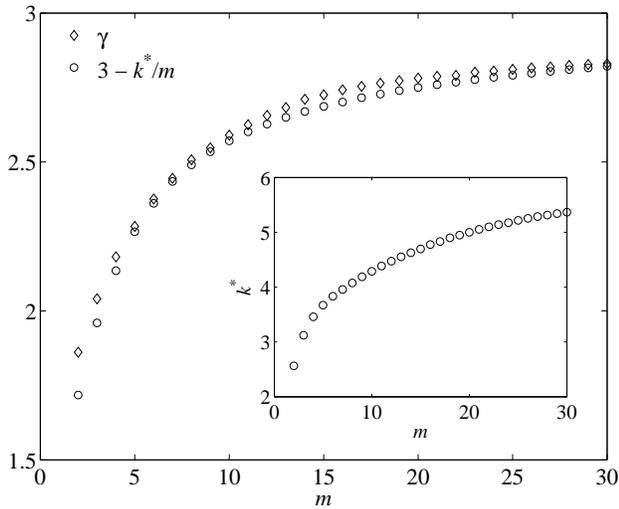}
  \caption{Local attachment procedure assigns effectively to each node an
    attractiveness proportional to $k_i-k^*$. The inset shows the offset $k^*$
    as a function of $m$. The diagram in the main figure shows the exponent
    $\gamma$ of the degree distribution and $(3-k^*/m)$.}
  \label{fig:k_star}
\end{figure}

Numerically we find that our local attachment growth model results in networks
with degree distributions that exhibit a power-law over a certain range of
degrees, see \figurename~\ref{fig:ba_local_degree_dist}. The exponential
cutoff of the degree distribution is a consequence of the finite size of the
network. The exponent $\gamma$ grows monotonically with parameter $m$, see the
inset of \figurename~\ref{fig:ba_local_degree_dist}.

As in the case of the fixed-size network model in
section~\ref{sec:Local_attachment} the origin of the power-law distribution of
the local attachment growth model can be understood by considering the mean
attractiveness. We find numerically that a node with degree $k\ge m$ attains
under the dynamics of local attachment an average attractiveness that is given
by the nonlinear function
\begin{equation}
  \label{eq:mean_A_growth}
  \langle A(k)\rangle = k-k^* \,.
\end{equation}
It can be shown analytically that network growth by preferential attachment
with an attractiveness $A(k)=k-k^*$ gives networks with a power-law degree
distribution with exponent $(3-k^*/m)$
\cite{Dorogovtsev:Mendes:Samukhin:2000:Structure_of_Growing_Networks}. For our
local attachment model the offset $k^*$ can be measured numerically. The value
$(3-k^*/m)$ and the empirical exponent $\gamma$ that is found by fitting the
degree distribution to a power-law agree very well, see
\figurename~\ref{fig:k_star}.

For local attachment networks without node removal the exponent $\gamma$ tends
to be smaller than for local attachment networks under churn. Node removal
makes it harder to gain a large number of neighbors and therefore the degree
distribution lacks nodes with very high degree in local attachment networks
under churn.

\begin{figure}
  \includegraphics[width=\linewidth]{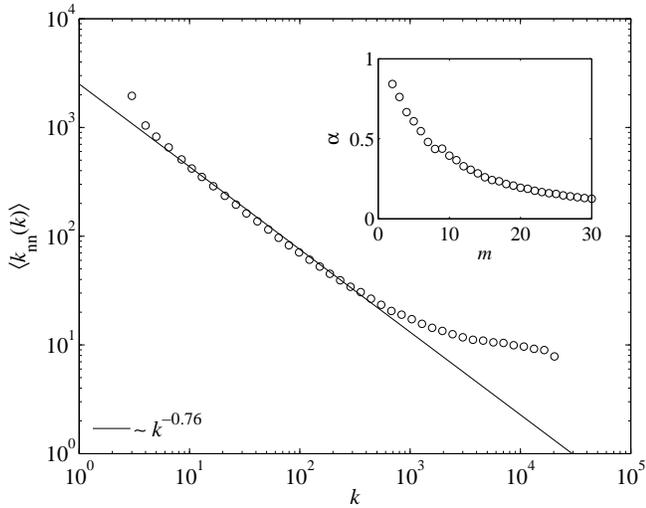}
  \caption{Distribution $\langle k_\mathrm{nn}(k)\rangle$ of the mean degree
    of the neighboring nodes of a node with degree $k$ for a growing local
    attachment network with $N=100\,000$ nodes and $m=3$. Inset: Exponent
    $\alpha$ of the power-law distribution $\langle
    k_\mathrm{nn}(k)\rangle\sim k^{-\alpha}$ of the mean degree of the
    neighbors of a node with degree $k$ as a function of the number $m$ of
    edges that connect a new node to the network.}
  \label{fig:local_attachment_alpha}
\end{figure}
\begin{figure}
  \centering
  \includegraphics[width=0.8\linewidth]{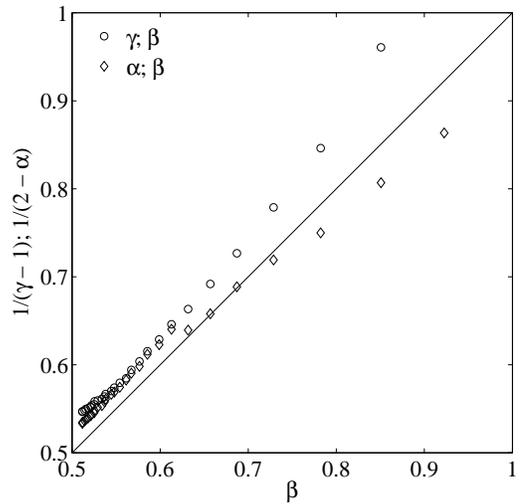}
  \caption{The scaling exponents $\alpha$ and $\beta$, and $\gamma$ and
    $\beta$ of growing local attachment networks satisfy approximately the
    relations $\beta=1/(\gamma-1)$, and $\beta=1/(2-\alpha)$
    respectively. Each exponent has been calculated by averaging over more
    than 100 networks with $N=100\,000$ nodes.}
  \label{fig:local_attachment_alpha_beta_gamma_scaling}
\end{figure}

\subsection{Scaling relations}

Local attachment growth networks can be characterized by further quantities
that follow a power-law distribution. The mean degree $\langle
k_\mathrm{nn}(k)\rangle$ of the neighboring nodes of a node with degree $k$
follows a power-law as well, see \figurename~\ref{fig:local_attachment_alpha}.
The exponent of this power-law $\langle k_\mathrm{nn}(k)\rangle\sim
k^{-\alpha}$ is negative and as a consequence nodes with large degree tend to
be connected to nodes having a small degree and vice versa. This can be
interpreted as a strong sign for disassortative correlations in the
probability distribution $p(k, k')$ that a randomly chosen edge connects two
nodes of degree $k$ and $k'$. Furthermore, $\alpha$ decreases with the
parameter $m$, that means the strength of the correlations depends on the
density of the network, see the inset of
\figurename~\ref{fig:local_attachment_alpha}.

The mean degree $\langle k(i, t)\rangle$ of a node that has entered the
network at time $t=i$ grows for $t\ge i$ following a power-law $\langle k(i,
t)\rangle\sim (t/i)^\beta$. Numerically we find that the exponents $\gamma$
and $\beta$ are approximately connected via the relation
\begin{equation}
  \label{eq:gamma_beta_growth}
  \beta = \frac{1}{\gamma-1}\,,
\end{equation}
and the exponents $\alpha$ and $\beta$ follow a scaling relation as well,
namely
\begin{equation}
  \label{eq:alpha_beta_growth}
  \beta = \frac{1}{2-\alpha} \,,
\end{equation}
see \figurename~\ref{fig:local_attachment_alpha_beta_gamma_scaling}.  In
\cite{Dorogovtsev:Mendes:2003:Evolution_of_networks} it is argued that the
scaling relations (\ref{eq:gamma_beta_growth}) and
(\ref{eq:alpha_beta_growth}) should be fulfilled by any growing network where
nodes of degree $k$ have an attractiveness $A(k)=k-k^*$.

Note that degree distributions $p(k)$ and $\langle k(i, t)\rangle$ are related
by the same scaling relations (\ref{eq:gamma_beta}) and
(\ref{eq:gamma_beta_growth}) in both variants of local attachment (nongrowing
and growing). But the character of $\langle k(i, t)\rangle$ is quite different
in the two cases, because for growing networks $\beta$ is the exponent of a
power-law, whereas for constant size networks under churn $\beta$
characterizes an exponential law.

\section{Conclusions}
\label{sec:Conclusions}

We have introduced local attachment as an universal node-joining protocol that
does not require global information. Nodes joining a network by local
attachment incorporate only \emph{local} information about the network
topology to produce stable \emph{global} properties. We investigated these
global properties of local attachment networks under churn as well as of
growing local attachment networks.

In both types of networks the degree distribution and other quantities follow
a power-law. The exponents of these power-laws fulfill simple scaling
relations. 

The local attachment procedure induces disassortative correlations between the
degrees of neighboring nodes and the emergence of the power-law degree
distribution for local attachment networks under churn is driven by these
correlations and by statistical fluctuations. 

Furthermore, correlations generate a nonlinear node attractiveness profile.
The qualitative features of local attachment networks can be described by an
effective model of preferential attachment that takes into account the
nonlinear node attractiveness profile.

\acknowledgments This work has been sponsored by the European Community's FP6
Information Society Technologies programme under contract IST-001935,
EVERGROW. We thank the Institute for Theoretical Physics at Magdeburg
University for providing computing time.

\bibliography{local-attachment}

\end{document}